**A Framework for Assessing Cumulative Exposure to Extreme Temperatures During Transit Trip**

**Huiying Fan**
School of Civil and Environmental Engineering
Georgia Institute of Technology
790 Atlantic Dr, Atlanta, GA 30332
Email: fizzyfan@gatech.edu

**Hongyu Lu**
School of Civil and Environmental Engineering
Georgia Institute of Technology
790 Atlantic Dr, Atlanta, GA 30332
Email: hlu305@gatech.edu

**Geyu Lyu**
School of City & Regional Planning
Georgia Institute of Technology
790 Atlantic Dr, Atlanta, GA 30332
Email: glyu8@gatech.edu

**Angshuman Guin, Ph.D.**
School of Civil and Environmental Engineering
Georgia Institute of Technology
790 Atlantic Dr, Atlanta, GA 30332
Email: angshuman.guin@ce.gatech.edu

**Randall Guensler, Ph.D.**
School of Civil and Environmental Engineering
Georgia Institute of Technology
790 Atlantic Dr, Atlanta, GA 30332
Email: randall.guensler@ce.gatech.edu

# A Framework for Assessing Cumulative Exposure to Extreme Temperatures During Transit Trip

**ABSTRACT**

The combined influence of urban heat islands, climate change, and extreme temperature events are increasingly impacting transit travelers, especially vulnerable populations such as older adults, people with disabilities, and those with chronic diseases. Previous studies have generally attempted to address this issue at either the micro- or macro-level, but each approach presents different limitations in modeling the impacts on transit trips. Other research proposes a meso-level approach to address some of these gaps, but the use of additive exposure calculation and spatial shortest path routing poses constraints meso-modeling accuracy. This study introduces HeatPath Analyzer, a framework to assess the exposure of transit riders to extreme temperatures, using TransitSim 4.0 to generate second-by-second spatio-temporal trip trajectories, the traveler activity profiles, and thermal comfort levels along the entire journey. The approach uses heat stress combines the standards proposed by the NWS and CDC to estimate cumulative exposure for transit riders, with specific parameters tailored to the elderly and people with disabilities. The framework assesses the influence of extreme heat and winter chill. A case study in Atlanta, GA, reveals that 10.2% of trips on an average summer weekday in 2019 were at risk of extreme heat. The results uncover exposure disparities across different transit trip mode segments, and across mitigation-based and adaptation-based strategies. While the mitigation-based strategy highlights high-exposure segments such as long ingress and egress, adaptation should be prioritized toward the middle or second half of the trip when a traveler is waiting for transit or transferring between routes. A comparison between the traditional additive approach and the dynamic approach presented also shows significant disparities, which, if overlooked, can mislead policy decisions.

# INTRODUCTION

With rapid urbanization and increasing frequency and severity of extreme temperature events, the thermal comfort of transit riders is becoming more important to the provision of safe and reliable daily travel. Extreme heat in the summer is a combined effect of extreme temperature and humidity coupled with the urban heat island effect (Kim, 2007), and is a common threat to urban citizens' health, particularly the elderly and people with special medical conditions (Fan et al., 2019). Wind chill, cold temperature coupled with winds that can steal body surface heat, is another extreme temperature event that is common in winter (Lin et al., 2019). Understanding exposure to extreme temperatures as travelers move throughout the transportation network is important for ensuring safe and comfortable travel.

While all travelers are affected by extreme temperatures, the level of impact across trips differs significantly. The use of transit modes typically exposes travelers to more outside conditions than the use of automobile. Underprivileged neighborhoods are particularly vulnerable to extreme temperatures due to higher reported exposure and lower access to mitigation options (Li et al., 2021; Nowak et al., 2022; Wilson, 2020). They also face lower automobile access, reducing their flexibility in modal choice (Klein & Smart, 2017). Senior citizens, who have increased susceptibility to extreme temperatures (Delclòs-Alió et al., 2019; Pantavou et al., 2013), use transit and sidewalks more frequently, again due to reduced flexibility in mode choice (Alves et al., 2020). Senior citizens also have reduced trip elasticity, or flexibility in making trip decisions (Amindeldar et al., 2017; Kruger & Drach, 2017), and are particularly vulnerable to extreme temperatures. Despite the need for individualized attention to specific groups who travel under extreme temperatures, current navigation apps have been designed primarily for "average" users. Weather condition, especially temperature, is not generally a factor in routing algorithms and is not provided as information in most available navigation systems (Apple, 2022; Google, 2022; MapQuest, 2022; Microsoft Bing, 2022; Nivetha et al., 2022).

**LITERATURE REVIEW**

In previous research, two general genres of studies have investigated the impacts of extreme temperatures on travelers, the "micro-level analysis" originates from urban and building design, wind dynamics, and biometeorology (Chatzidimitriou & Yannas, 2016; Jamei & Rajagopalan, 2018; Labdaoui et al., 2021; D. Li et al., 2021; Taleghani et al., 2016), and the "macro-level analysis" focuses on transportation network properties and user behaviors (Guo, 2009; Guo & Loo, 2013; H. Kim, 2015; Peiravian et al., 2014).

Microscale analyses are often conducted at the scale of the building, street section, or city block. They often focus on the benefits of urban features in reducing exposure, such as green roofs and pavements (Taleghani et al., 2016), courtyards (Chatzidimitriou & Yannas, 2016), aspect ratio, canyon effect, and lightning (Jamei & Rajagopalan, 2018; Labdaoui et al., 2021). This type of research often applies fine-grained biometeorological measures like the physiologically equivalent temperature (PET) (Nouri et al., 2018), and simulation methods, including the ENVI-met (Jamei & Rajagopalan, 2018) and the RayMan model (Matzarakis et al., 2007, 2010). Microscale studies bring high granularity and tangible outputs but are often limited to a constrained area due to the high computational demand.

Macroscale analyses focus on traveler behavior and network characteristics such as safety (Guo & Loo, 2013), walkability (Guo, 2009), environmental quality (Peiravian et al., 2014), and urban greenery (H. Kim, 2015). Macroscale analyses typically rely more on qualitative measurements, including a qualitative scoring system (H. Kim, 2015; Peiravian et al., 2014) and revealed preference modeling (Guo, 2009; Guo & Loo, 2013). Macroscale studies can provide holistic consideration of travel patterns over an entire trip but often do not provide much support for systems engineering design and optimization. Few, if any, macro-level analyses have focused explicitly on temperature and pedestrian thermal comfort.

Recent studies have begun integrating micro- and macro-level approaches to yield insights that are both tangible and comprehensive. Y. Liu et al. (2024) merged trip trajectories with high-resolution temperature data to provide trip-level quantifiable outputs. However, their model's reliance on a consistent activity and exposure profile, typical of fixed-pace jogging, limits its applicability to a typical trip. Li et al. (2023) advanced the field by integrating a Work/Rest schedule to calculate cumulative exposure. However, their use of a Dijkstra-based algorithm did not fully capture all transit trip segments, such as wait times. Huang et al. (2024) and L. Liu et al. (2024) approach offered insights into the spatio-temporal interactions among transit mode segments but are constrained by an additive exposure calculation, which may not fully reflect realistic physiological responses.

To understand the complex mechanisms of cumulative heat exposure during transit trips, this research presents a method that integrates the dynamics of both heat stress accumulation and the spatio-temporal transit profile. Operating at the meso-scale level, the approach similarly bridges the gaps between micro-scale and macroscale methods presented in prior research.

**METHODOLOGY**

This section outlines the data sources and methodologies employed in the study, illustrated as a schematic in Figure 1. First, the thermal comfort analytic module gathers meteorological data and establishes thermal comfort standards using the National Weather Service (NWS)'s equations. The transit network used by TransitSim 4,0 is developed for each modeling time period using GTFS data for routes, stops, and schedules. Modeled transit trips are then processed with a travel conditions module to generate second-by-second trajectories for each trip. This modeling also uses the Compendium of Physical Activities to quantify the intensity levels associated with travel activity in each trip segment (e.g., walk to transit, wait for bus, etc.) for different populations (average adults, older adults, and wheelchair users). Cumulative exposure is then calculated using NIOSH's Work/Rest schedule, using the second-by-second physical activity and thermal comfort profiles to calculate cumulative exposure and trip-level extreme heat risk. Finally, the resilience of the transit system is assessed by evaluating the criticality, exposure, and vulnerability of network elements. Two prioritization scores are developed: a mitigation-based score that assesses risk accumulation before potential heat-related illnesses, and an adaptation-based score that offers insights into immediate medical support needed post-exposure. Detailed descriptions of each module follow in the subsequent sections.

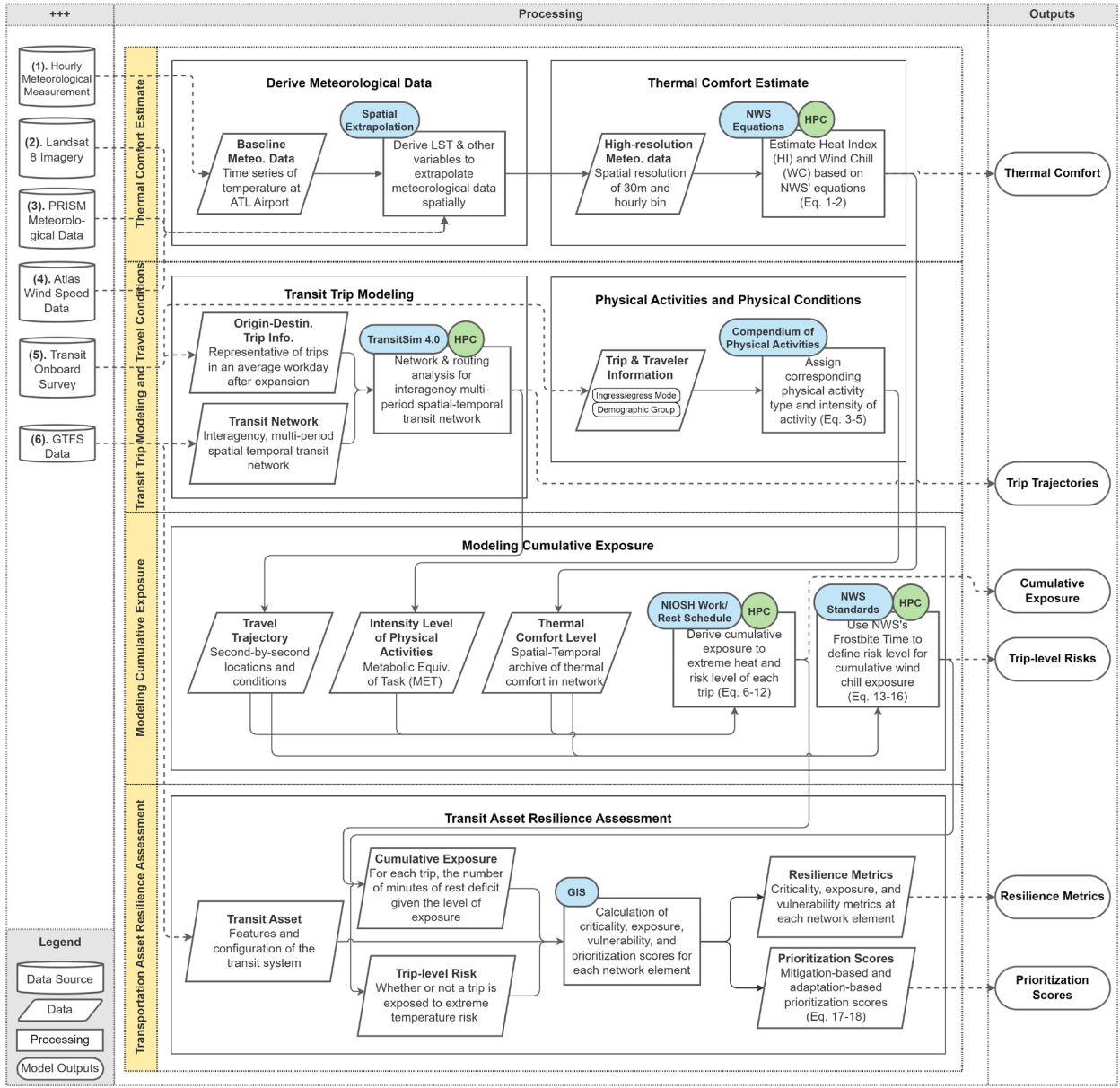

**Figure 1.** Overview of Analytical Framework

**Thermal Comfort Estimates**

This section details methods for temperature calculation and the derivation of thermal comfort indices. High-resolution (both spatially and temporally) temperature data are required to assess each traveler's exposure at their specific travel time and location. While gauge data typically provide high temporal resolution, they are often only sparsely available in a city. The HeatPath Analyzer framework uses the hourly binned air temperature at the center of the study area as the basis of temperature estimates. Spatial variation in air temperatures at the intra-city level can create significant disparities within the study area, with differences up to 7.6 degrees Celsius due to variations in land cover(Cao et al., 2021). In this research, spatial variation is extrapolated using land surface temperature. Details of the retrieval methods can be found in Appendix A. The output of the temperature calculation module is a dataset of temperature derivation for every 30m by 30m location grid and every one-hour time point.

Thermal comfort can be estimated using derived temperature data. Researchers have developed over 150 thermal stress indicators over the past century (Budd, 2008; Chatzidimitriou & Yannas, 2016; Höppe, 1999; Ioannou et al., 2022; Matzarakis et al., 2007, 2010; Nouri et al., 2018; Taleghani et al., 2016; Thorsson et al., 2014). In this study, the traditional National Weather Service (NWS) heat index (Rothfusz & Headquarters, 1990) and wind chill (NWS, 1992) serve as indices for thermal comfort, for their solid empirical foundation, wide application in research, easily obtainable inputs, and reasonable assumptions representative of the general U.S. population.

The heat index is calculated by temperature and relative humidity, and the wind chill is calculated by temperature and wind speed. The relative humidity data is calculated from the PRISM Climate Group data (PRISM Climate Group, 2014), a widely recognized and adopted dataset for historical weather conditions, maintained by Oregon State University. Wind speed data are obtained from the Global Wind Atlas 3.0, a free, web-based application developed by the Technical University of Denmark (Denmark Technical University & World Bank Group, 2015).

The heat index ($T_{HI}$) is calculated using Equation 1 (Rothfusz & Headquarters, 1990):

$$T_{HI} = -42.379 + 2.04901523 \times T + 10.14333127 \times RH$$
$$- 0.22475541 \times T \times RH - 0.00683783 \times T^2$$
$$- 0.05481717 \times RH^2 + 0.00122874 \times T^2 \times RH \quad (1)$$
$$+ 0.00085282 \times T \times RH^2 - 0.00000199 \times T^2 \times RH^2$$

Where $T$ is the derived temperature and $RH$ is the derived relative humidity. Adjustment factors are applied for specific situations as outlined by the National Weather Service (NWS, 2022). The wind chill ($T_{WC}$) is calculated with temperature and wind speed ($V$) (Equation 2) as outlined by the National Weather Service (NWS, 1992).

$$T_{WC} = 35.74 + 0.6215 \times T - 35.75 \times V^{0.16} + 0.4275 \times T \times V^{0.16} \quad (2)$$

The calculated heat index ($T_{HI}$) and wind chill ($T_{WC}$) are both in Fahrenheit, approximating the apparent temperature (the temperature that feels like to human bodies under given temperature conditions combined with relative humidity and wind speed).

**Transit Trip Modeling and Travel Conditions**

Transit modeling is conducted using the open-source General Transit Feeds Specification (GTFS) data (Google Developers, 2024) along with TransitSim 4.0, a model developed, maintained, and regularly updated by the NCST research team at the Georgia Institute of Technology (Fan et al., 2022; Li, 2019; Li et al., 2018). TransitSim 4.0 includes interagency modeling capacity and incorporation of the RAPTOR algorithm for efficient transfer and wait time simulation. This allows it to provide precise spatio-temporal information and generate second-by-second locations for each transit rider (Fan, Lyu, Guin, et al., 2024). The transit trip modeling outputs a set of second-by-second trip trajectories for each sample, detailing specific mode changes that facilitate the analysis of exposure across different modes when combined with activity profiles.

Under identical meteorological conditions, the human body's response to extreme heat can vary significantly across different populations and physical activity intensities, due to differences in metabolic rates (Kenney & Munce, 2003; Sawka et al., 2011). A standardized measure is needed to quantify the metabolic rate associated with different physical activities for various populations. The Compendium of Physical Activities (Compendium), developed, maintained, and continuously updated by four U.S. National Institute of Health-funded research centers across universities for over 30 years, is designed to provide consistent assignment of physical activity intensity levels associated with their health outcomes. It has accumulated over 22,000 combined citations (Ainsworth et al., 2024), providing separate measures for average adults (Herrmann et al., 2024), older adults (Willis et al., 2024), and wheelchair users (Conger et al., 2024). The Compendium uses the Metabolic Equivalent of Task (MET) to reflect the individual energy cost of activities (MET60+ for older adults). For this study, intensity levels of physical activities ($W_{level}$) are based on their MET value ($MET$) and whether the activity is air-conditioned ($C$). The categorization is as follows (Eq. 3):

$$W_{level}(C, MET) = \begin{cases} \text{"rest"}, & C = \text{"yes"} \\ \text{"light"}, & C = \text{"no" and } M \leq 1.5 \\ \text{"moderate"}, & C = \text{"no" and } 1.5 < M < 4 \\ \text{"heavy"}, & C = \text{"no" and } M \geq 4 \end{cases} \quad (3)$$

Where MET value ($M$) is defined by demographic group ($demo(i)$) of individual $i$ and their Physical Activity ($pa(t)$) at period $n$:

$$MET(t) = f(demo(i), pa(n)) \quad (4)$$

Table 1 summarizes the activities involved in each mode, along with their corresponding MET ($MET_{60+}$) values and intensity levels.

**Table 1.** MET Values and Intensity Levels for Different Physical Activities across Populations

| Mode | Activity | Conditioned | Average Adults | | Older Adults | | Wheelchair Users | |
|---|---|---|---|---|---|---|---|---|
| | | | MET | Intensity | MET60+ | Intensity | MET | Intensity |
| Automobile | Automobile driving | Yes | 2.0 | Rest | 2.3* | Rest | 2.0* | Rest |
| | Automobile riding | Yes | 1.3 | | 1.5* | | 1.3 | |
| Transit | Transit riding | Yes | 1.3 | Rest | 1.5* | Rest | 1.3 | Rest |
| Wait | Sitting quietly | No | 1.0 | Light | 1.3 | Light | 1.0* | Light |
| | Standing quietly | No | 1.3 | | 1.5 | | 1.3* | |
| Walk | Walking for transportation (level, moderate pace) | No | 3.5 | Moderate | 4.3 | Heavy | | |
| | Walking uphill (1-5% grade, moderate pace) | No | 5.3 | | 5.0 | | | |
| | Walking uphill (6-10% grade, moderate pace) | No | 7.0 | | 7.3* | | | |
| Bike | Bicycling (to/from work) | No | 6.8 | Heavy | 5.3 | Heavy | | |
| Micro-mobility | Motor scooter | No | 2.8 | Moderate | 3.1* | Moderate | | |
| | Motorcycle | No | 2.8 | | 3.1* | | | |
| Wheelchair | Wheeling on sidewalk | No | | | 4.0* | Heavy | 3.2 | Moderate |

* Estimated based on average adults' values

Based on Table 1, equation 4 can be aggregated to:

$$MET(t) = f(demo(i), M(n)) \qquad (5)$$

Where $M(n)$ is the mode of transportation for the period $n$.

**Modeling Cumulative Exposure**

With second-by-second thermal comfort and intensities of physical activities estimated, an equation is needed to establish the relationship between thermal exposure patterns and physiological response among individual pedestrians.

*Cumulative Heat Exposure*

Our study employs modeling frameworks used in occupational safety, recognized for their focus on individual-level risk estimates and their robust, validated methodologies. We adapted the Work/Rest schedule proposed by the U.S. National Institute for Occupational Safety and Health (NIOSH) (Centers for Disease Control et al., 2017) to calculate both the cumulative heat exposure $E_{HI}(i,j)$ and associated risk factor $R_{HI}(i,j)$ for individual $i$ throughout trip $j$. A trip $j$ is the comprehensive travel trajectory from origin to destination. The Work/Rest schedule is composed of a series of periods. $N_j = \{1,2,3,\ldots,n_j\}$, their corresponding travel time $T_j = \{t_1, t_2, t_3, \ldots, t_{n_j}\}$, mode $M_j = \{m_1, m_2, m_3, \ldots, m_{n_j}\}$, and geospatial location $(X_j, Y_j) = \{(x_1, y_1), (x_2, y_2), \ldots, (x_{n_j}, y_{n_j})\}$.

At any period $n \in N_j$, the intensity level of individual $i$'s physical activity $W_{level}(i, n)$ can be found by combining equations (3) - (5):

$$W_{level}(i, n) = W_{level}(C_{m_n}, M(i, n)) \tag{6}$$

We adapted the work/rest schedule proposed by NIOSH (Centers for Disease Control., 2017) to find the maximum appropriate continuous interval for physical activity ($\rho_{i,n}$) and the corresponding length of rest needed ($\eta_{i,n}$), both defined as time in minutes:

$$(\rho_{i,n}, \eta_{i,n}) = H(T_{HI}(x_n, y_n), W_{level}(i, n))$$
$$= \begin{cases} (1 \times 10^4, 0), & \text{if } T_{HI} < 90 \\ (1 \times 10^{-6}, 1 \times 10^{-5}), & \text{if } T_{HI} > 112 \\ Lookup(T_{HI}, W_{level}), & \text{if } 90 \leq T_{HI} \leq 112 \end{cases} \tag{7}$$

At period $n$, a heat risk flag ($r_{HI}(i, n)$) is triggered when the cumulative amount of physical activities burden ($P_{i,n}$) exceeds 1 or when ($P_{i,n}$) reaches one yet rest is not provided in the next period:

$$r_{HI}(i, n) = 1_{\{P_{i,n} > 1\} \cup \{P_{i,n} = 1 \wedge W_{level}(i, n+1) = \text{"rest"}\}} \tag{8}$$

Where $P_{i,n}$ at period $n$ is defined as the cumulative physical activities ($p_{i,k}$) burden summing across the current and all previous periods, where $p_{i,k}$ is defined as the ratio of travel time at period $k$ and the maximum appropriate continuous travel internal $\rho_{i,k}$:

$$P_{i,n} = \sum_{k=1}^{n} p_{i,k} = \sum_{k=1}^{n} \frac{t_k}{\rho_{i,k}} \tag{9}$$

The trip-level extreme heat risk factor $R_{HI}(i, j)$ is thus:

$$R_{HI}(i, j) = 1_{\{\exists n \in N_j | P_{i,n} = 1\}} \tag{10}$$

In this study, the incremental heat exposure at period $n$ ($e_{i,n}$), when period $n$ is not a rest period, is expressed as time in minutes, representing the total amount of time needed for one to fully recover from the effect of heat exposure. When period $n$ is a rest period, the incremental heat exposure represents the amount of rest taken and resulting in reduced heat stress:

$$e_{i,n} = \frac{\eta_{i,n} \times t_n}{\rho_{i,n}} \cdot 1_{\{W_{level}(i,n) \neq \text{rest}\}} - t_n \cdot 1_{\{W_{level}(i,n) = \text{rest}\}} \tag{11}$$

The cumulative heat exposure $E_{HI}(i,j)$ over a trip $j$ is the sum of all incremental heat exposure throughout the periods:

$$E_{HI}(i,j) = \sum_{k=1}^{n_j} e_{i,k} \tag{12}$$

*Cumulative Wind Chill Exposure*

The wind chill exposure is defined based on frostbite time defined by the U.S. National Weather Service (NWS, 1992):

$$\tau_n = \Gamma(T_{WC}(x_n, y_n)) \tag{13}$$

In this study, incremental wind chill exposure at period $n$ ($\varepsilon_n$) is the proportional contribution of the exposed time to frostbite probabilities:

$$\varepsilon_n = \frac{t_n}{\tau_n} \tag{14}$$

The cumulative wind chill exposure $E_{WC}(j)$ throughout the entire trip is:

$$E_{WC}(j) = \sum_{k=1}^{n_j} \varepsilon_k = \sum_{k=1}^{n_j} \frac{t_k}{\tau_k} \tag{15}$$

The trip is flagged for wind chill risk when the cumulative wind chill exposure is larger or equal to 1, or when the cumulative exposure reaches frostbite time:

$$R_{WC}(j) = 1_{\{E_{WC}(j) \geq 1\}} \tag{16}$$

**Transit Asset Resilience Assessment**

To inform future transit planning, the effects of extreme temperatures from individual trips are consolidated into a composite assessment to evaluate their impact on the entire transit asset system. Given the focus of our investigation, where in-vehicle periods are air-conditioned and thus generally pose no additional health risks, our analysis primarily concentrates on stations and ingress, egress, and transfer footpaths (including all active last-mile transportation modes such as wheelchair, bike, and micromobility). Let $A$ denote the overall asset network, then $A = A_S \cup A_F$, where $A_S = \{a_{s,1}, a_{s,2}, a_{s,1}, \ldots, a_{s,l}\}$ and $A_F = \{a_{f,1}, a_{f,2}, a_{f,1}, \ldots, a_{f,m}\}$ represent stations and footpaths, respectively.

In transportation asset management, scholars have proposed different metrics to represent the multifaceted impacts of extreme weather on the resilience of the network. This study, uses three measures: (1). Exposure, the frequency, and severity of disturbances. In this study, exposure is operationalized as the average amount of rest required per minute spent on the infrastructure, given the temperature and humidity conditions; (2). Vulnerability, the susceptibility of an asset or system to extreme temperatures. In this study, vulnerability is the percent of incidents on each asset segment that experience elevated risk (Fan et al., 2023); and (3). Criticality, is the importance of a particular asset element within the broader network (Meerow et al., 2016; Ribeiro & Gonçalves, 2019). In this study, criticality is defined as the total travel time across all travelers, spent on a particular asset (i.e., per station, or specific unit length of sidewalk).

The analysis yields two prioritization scores. For a transit asset system composed of stations and footpaths, the Mitigation-based Prioritization Score for an asset segment $a_o \in A$, denoted as $s_{mitigation}(a_o)$, is defined by the product of exposure and criticality. This score quantifies the total contribution of risk each segment induces for all trips (Eq. 17). The Adaptation-based Prioritization Score at asset segment $a_o \in A$, denoted as, $s_{adaptation}(a_o)$, is calculated as the product of vulnerability and criticality per unit of travel time. This score represents the total outcome of risk that travelers experience in each segment (Eq. 18).

$$s_{mitigation}(a_o) = f_{exposure}(a_o) \times \sum_{j \in J} \sum_{n \in N_j} t_n \cdot \delta_n^{a_o} \qquad (17)$$

$$s_{adaptation}(a_o) = f_{vulnerability}(a_o) \times \sum_{j \in J} \sum_{n \in N_j} \delta_n^{a_o} \qquad (18)$$

**CASE STUDY**

A case study for the Atlanta Metropolitan Area ("Metro Atlanta") in Georgia illustrates the functionality of the HeatPath Analyzer model and demonstrates the mechanisms of cumulative heat exposure during transit trips. As of 2020, metro Atlanta was home to more than six million people. Despite a focus on freight transportation and automobile transportation (Crimmins & Preston, 1980; Henderson, 2002), metro Atlanta has established a mature transit system with various options for riders, including fixed-routes transit (rails, buses, express buses), circulators and shuttles, and on-demand services (ARC, 2021). Public transportation in metro Atlanta is operated by multiple transit agencies (i.e., transit service providers). This case study focuses on the fixed-route transit options in five transit agencies (MARTA, Atlanta Streetcar, GRTA Express Bus, Gwinnett County Transit, and CobbLinc) providing service in the study area (see Figure 2).

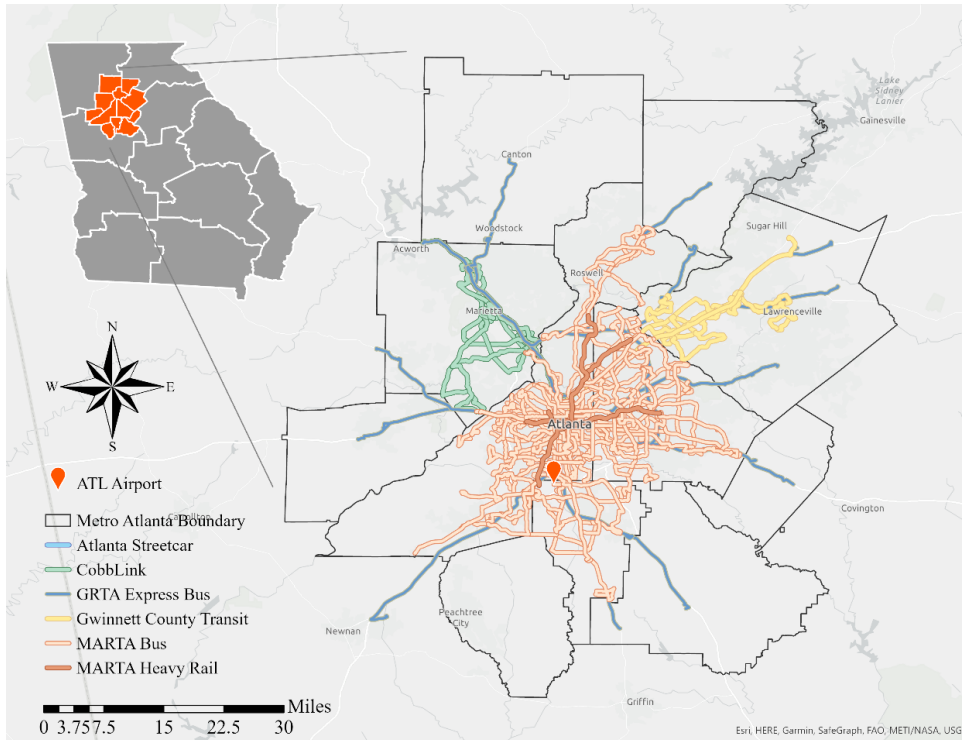

**Figure 2.** Study Area, Transit Agencies and Routes, and Location of ATL Airport
(example network configuration during March 7th to April 18th, 2019)

The origin-destination and trip start time information used in the analyses is taken from the ARC Transit Onboard Survey, which collected 43,398 transit rider interview responses, over multiple routes and over multiple days from March through December 2019 (ARC, 2020). The ARC onboard survey represents the most comprehensive transit survey with demographic information obtained for transit riders before the pandemic (and represents a sample of about 10% of the region's daily ridership). Each trip in the survey is recorded as a separate sample. The survey methodology also includes an expansion process where responses are weighted and expanded to the metro area using several key factors: route, direction, time-of-day, and route segment. After this weighting and expansion, the dataset encompasses a total of 211,057 trip-equivalents, which are representative of both the volume and composition of trips during an average weekday across the region (Atlanta Regional Commission, 2019). Samples in the original data set with incomplete information (1,756) were removed from the analyses, leaving 40,289 records (95.8% of the original data). Each trip record includes a variety of demographic variables, such as gender, age group, and whether the individual is a wheelchair user. In addition to trip origin-destination information, each sample also contains trip-specific information used in calculating the physical intensity of travel activities (such as departure time, transit access mode, route transfers, etc.).

The longitudinal onboard transit survey covers a 10-month study period, during which the transit network and service charactersitcs were updated 11 by the service providers, resulting in 12 different networks during the study period. There are some notable differences across these transit network periods. Therefore, separate network construction and trip routing are conducted for each modeling period. Given the high computational demand required in the study, the Python-based analysis is launched on the PACE High-Performance Computing (HPC) framework at Georgia Tech (Georgia Institute of Technology, 2022).

**RESULTS**

This section presents the results of the analyses, beginning with an overview of thermal comfort estimates, followed by a detailed inspection across modes and trajectories. The next section summarizes the output of the transit network resilience assessments. Finally, a section compares the modeling outputs from the existing approach and approaches should the estimation be conducted in an additive or solely spatial manner..

**Overview of Exposure to Extreme Temperature**

After weighting and expansion, the analytical dataset represents 198,416 trips on an average weekday during the study period. The median travel time among all trips is 71.88 minutes. The survey does not include samples from June or July of 2019, because the school summer break makes these samples non-representative of average weekdays. For the survey period of April to December 2019, Atlanta had a long summer from the second half of May to the end of September. In summer, the highest temperatures were typically observed in the city center, along major transportation corridors (I-75 and I-85), and in built-up areas in nearby counties. In terms of time, the highest temperatures typically occurred between 10:00 A.M. and 9:00 P.M., with the peak typically occurring between 1:00 P.M. to 5:00 P.M. The winter season observed during the study period is relatively shorter, including November and December 2019. In winter, the lowest temperature is most often observed in the northwest or southeast regions of the city, and Built-up areas often experience higher temperatures than other places. While afternoons (1:00 P.M. to 4:00 P.M.) are typically warmer than other periods, cold winter days often see consistent cold temperatures throughout the entire day. A detailed view of the temperature during the study period can be found in Appendix B.

In the case study, 5,082 (2.6%) of the total 198,416 trips are exposed to extreme heat risk, and 0 (0.0%) are exposed to wind chill risk. The average magnitude of exposure to extreme heat quantified as the number of minutes needed for the human body to rest to fully restore to pre-exposure status ("rest deficit"), is 0.73 minutes, with a large variance across samples (standard deviation 13.08 minutes). Median trips at risk see 3.23 minutes of rest deficit, where the mean is 24.81 minutes with a large variance (standard deviation of 77.79 minutes). Wind chill rest deficit, on the other hand, has a very small average ($7.5 \times 10^{-4}$ minutes) and a small variance ($3.5 \times 10^{-4}$ minutes).

**Figure 3** illustrates the percentage of trips exposed to extreme heat by month, the distribution of these trips across different times of the day, and the geographical distribution of the travelers' home locations (aggregated to TAZ centroids to protect personally identifiable information). All trips at risk of extreme heat occur in the late summer and mid-fall, specifically from August to October. Notably, June and July were excluded from this analysis for school holidays. During the analyzed months, the percentages of daily trips exposed to extreme heat risks were 10.2% in August, 6.45% in September, and 2.7% in October. These percentages translate to approximately 20,238, 12,797, and 5,437 transit trips, respectively, on an average weekday in the summer of 2019. Such levels of exposure significantly surpass the occupational safety standards set by NIOSH.

The majority of trips at risk of extreme heat were predominantly concentrated in the afternoon, between 2 P.M. and 4 P.M., with additional occurrences extending into the late morning starting at 10:00 A.M., and a few extending into the early evening until 7:00 P.M. The home locations of travelers most susceptible to extreme heat are primarily in Downtown and Midtown Atlanta, southern Atlanta, and along the I-85 interstate corridor.

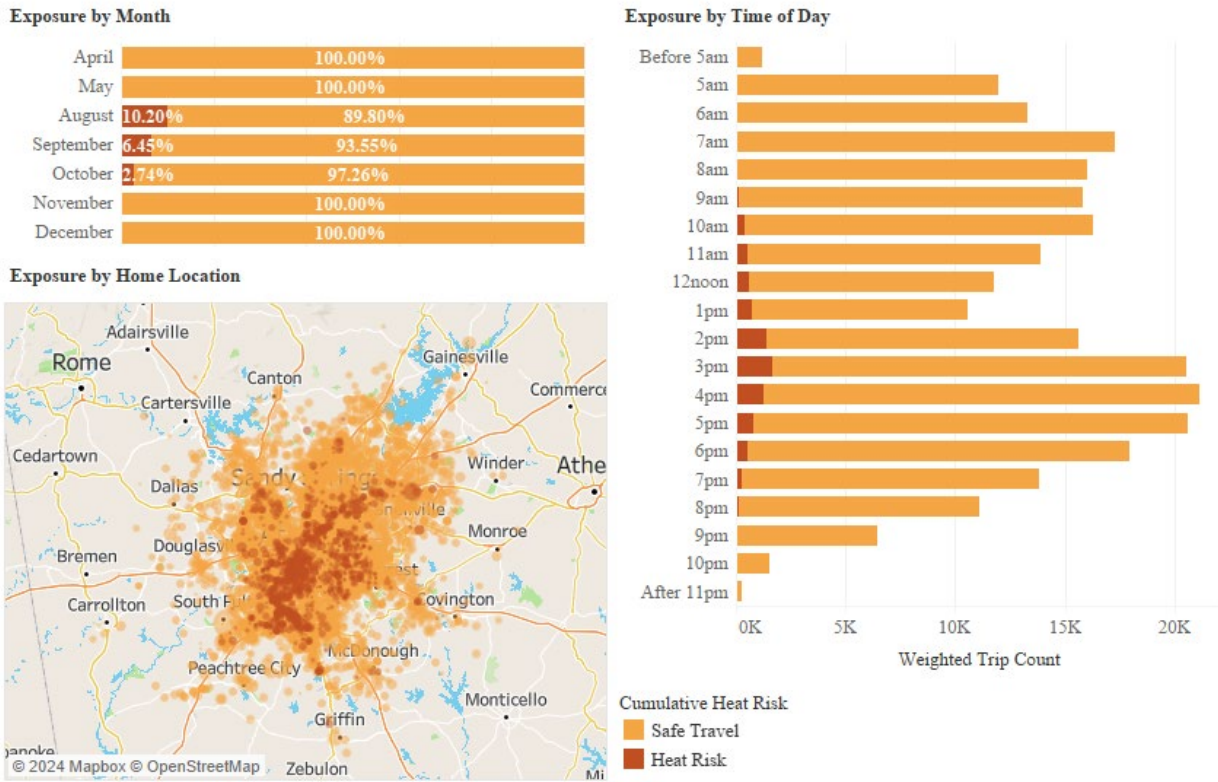

**Figure 3.** Number of Trips Exposed to Extreme Heat by Month, Time of Day, and Home Location

**Example of Cumulative Trip-Level Heat Exposure**

The HeatPath Analyzer generates second-by-second activity profiles and thermal comfort levels for each trip, laying the groundwork for an in-depth analysis of heat exposure mechanisms across different transit mode segments.

Figure 4 illustrates such a profile using an example trip on August 14, 2019. Departing at 2:43 P.M., the individual began their trip by biking for 1.2 minutes, at an average Heat Index of 96.3 ºF. With biking categorized as high-intensity physical activity, this leg of the trip generated 0.39 minutes of rest deficit but did not trigger a health risk. This was followed by a 15-minute wait at a bus station at 92.5 ºF with low physical intensity, and then a 17-minute bus ride in an air-conditioned vehicle. Shortly after boarding the bus, the individual's energy was restored, and the rest deficit dropped to zero. Arriving at the first transfer location at 3:17 P.M., the traveler began a walking transfer for 5.4 minutes at an average Heat Index of 104.0 ºF and moderate physical intensity. The first half of the walk transfer was at a higher temperature, thus accumulating more demand for rest. This leg accumulated a total of 11.1 minutes of rest deficit, with a risk flag triggered. After that, the traveler waited for another 10.7 minutes at light physical intensity without additional exposure accumulation. At 3:33 P.M., the traveler boarded the second bus, riding for 22.0 minutes in an air-conditioned environment and fully restoring their physical status. The final leg of their journey was a 0.8-minute walk to reach their destination. This trip is considered at risk of extreme heat exposure because the traveler faces continued exposure for 26.8 minutes after a health alert was triggered.

Profiles like this capture detailed second-by-second data on location, thermal comfort, and physical activity, and are generated for every trip. By aggregating these data across all observed trips, we synthesize a holistic view of travel conditions and cumulative heat exposure.

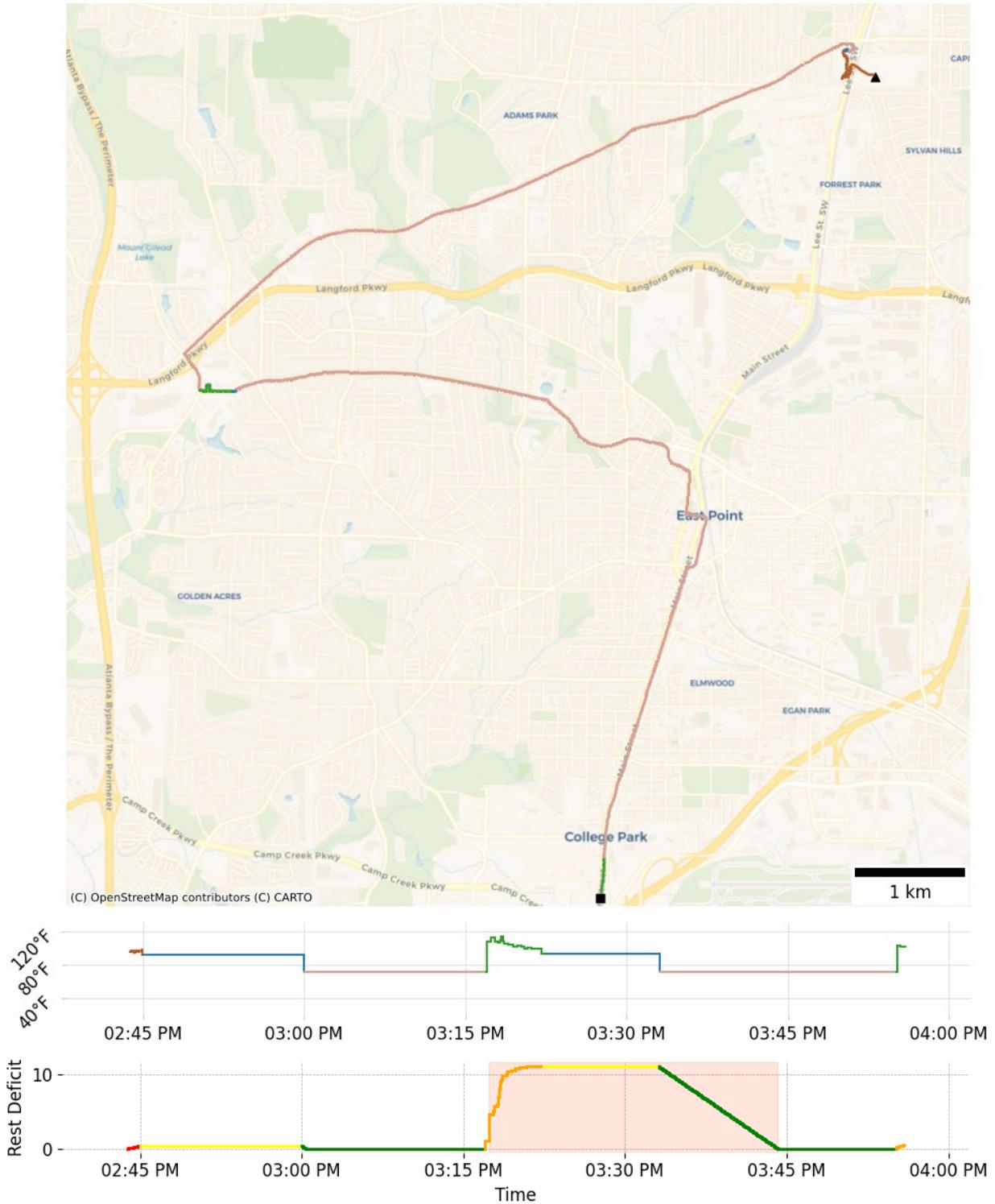

**Figure 4.** Example of a Modeled Travel Trajectory (**Mode colors** – brown: bike, blue: wait, green: walk, pink: transit; **Intensity colors** – red: high intensity; orange: moderate intensity; yellow: low intensity; green: rest; coral background: extreme heat flag triggered)

**Exposure by Transit Mode Segments**

The vast majority of the total 198,416 modeled trips (198,364, 99.97%) include one or more waiting events, usually riders waiting for the next transit vehicle at a route transfer or transit ingress location. A total of 362,818 wait events were modeled (an average of 1.83 wait events per trip). Concerning transit trip access, the majority of trips include a walk ingress (152,615, 76.9%) and egress (155,590, 78.4%), followed by automobile ingress (34,339, 17.3%) and egress (32,179, 16.2%), see Figure 5.

Out of the 71.88-minute average trip-level travel time, waiting periods (combining ingress and transfer) accounted for the largest time segment, averaging 34.16 minutes per trip, followed by transit riding at 18.17 minutes. When a wheelchair was used for egress, the average travel time was notably longer, at 23.14 minutes. Micromobility also had a significant travel time, averaging 13.13 minutes for ingress and 7.91 minutes for egress. Walking for ingress, egress, and transfer also represented a substantial duration, averaging 10.08, 9.99, and 6.92 minutes respectively. Automobile ingress and egress averaged 8.45 and 8.9 minutes, while Bike ingress and egress times were 5.92 and 6.01 minutes respectively.

The total heat exposure (mitigation-based) across all trips, quantifying the total rest deficit a mode segment contributes to the trip, is the largest in walk trips. Specifically, walk ingress and egress account for 10,183 and 9,690 minutes, respectively. Waiting segments are the next most significant, contributing 9,317 minutes, followed by walk transfers at 2,172 minutes. All other modes, contributing less than 100 minutes each, are considered negligible and are not detailed further. In contrast, the pattern of wind chill exposure differs from that of extreme heat. Waiting segments pose the highest risk, totaling 959.7 minutes, with ingress, egress, and transfers contributing 222.6, 209.9, and 133.8 minutes, respectively.

Total heat vulnerability (adaptation-based), quantified by the number of trips experiencing extreme heat risk per mode, exhibits a pattern markedly different from the mitigation-based exposure. Notably, all in-transit modes display high vulnerability, with walk transfers experiencing the most, affecting 1,140 trips. This is followed by waiting segments with 1,045 trips, and transit riding with 945 trips. A notable disparity is observed between walk egress and walk ingress, where egress records significantly higher vulnerability at 633 trips compared to only 199 trips for ingress. Wind chill does not pose any risk on trips.

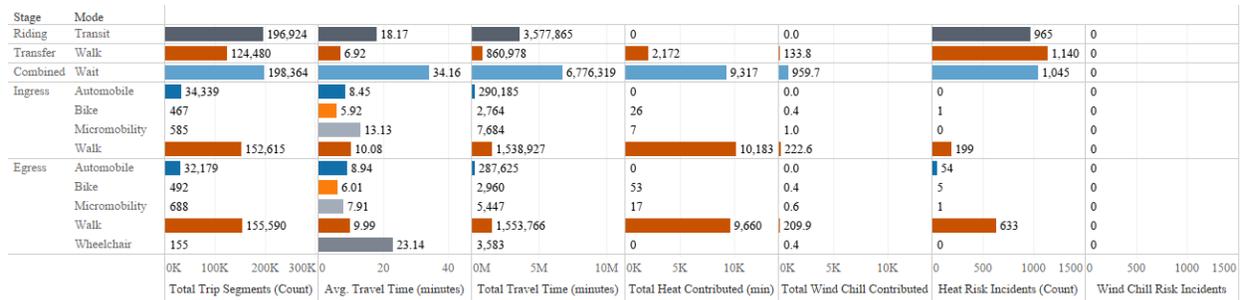

**Figure 5.** Travel and Exposure Metrics Breakdown by Mode

**Transit Network Resilience to Extreme Temperatures – Recommendations for Prioritization**

As defined in the methods, criticality for each station (Eq. 17) quantifies the total wait time experienced by all travelers throughout all trips on an average weekday. In Atlanta, stations with the greatest criticality are located near downtown Atlanta (156,698 hours), followed by Airport Station (90,837 hours), and Lindbergh Center Station (82,213 hours). Peripheral stations such as those near Marietta, Lawrenceville, and Clayton County exhibit medium criticality. Criticality for footpaths is measured by total transfer time on a per-mile basis. The footpaths with the highest criticality are located around Atlanta Airport, Midtown Atlanta, and areas near Buckhead, Sandy Springs, and Lawrenceville (Figure 10 in Appendix C).

Figure 6 shows exposure, vulnerability, and prioritization scores for extreme heat. Due to its minimal influence in the study area, wind chill is not discussed in detail in the main text, but can be found in Figure 11 of Appendix D. Figures 6a and e illustrate the average heat exposure per station and footpath, measured as rest deficit each network elements contribute on average. Figures 6c and g present the average vulnerability scores, quantified as the percentage of incidents traveling through the network element that encounters heat risk. High-exposure areas are near major transportation corridors and in dense urban areas with a high percentage of impervious surfaces. Conversely, vulnerability scores are more dispersed.

Figure 6b, d, f, and h depict the overall prioritization scores, which are composite metrics of exposure, vulnerability, and criticality. Figure 6b and f highlight the mitigation-based prioritization scores (combining exposure and criticality, Eq. 17), whereas Figure 6d and h detail the adaptation-based scores (combining vulnerability and criticality, Eq. 18). These prioritization scores reveal distinctly different focal points from those of exposure and vulnerability alone, with a strong inclination towards high-criticality locations. Notably, the mitigation-based heat prioritization score is highest around the Airport, Dawson, and Southlake areas. In contrast, the adaptation-based prioritization score emphasizes Doraville central station and its surrounding footpaths, Cumberland Station, Southlake Mall Station, and Lenox Square.

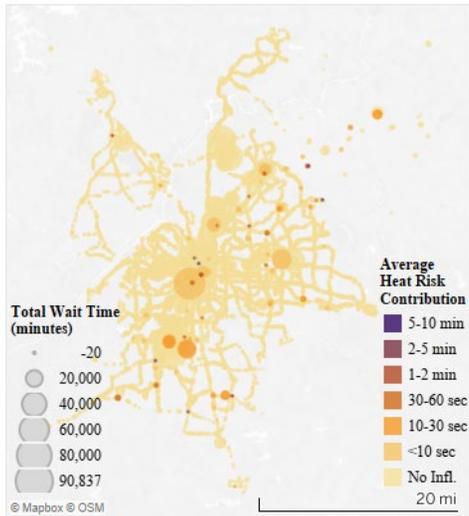 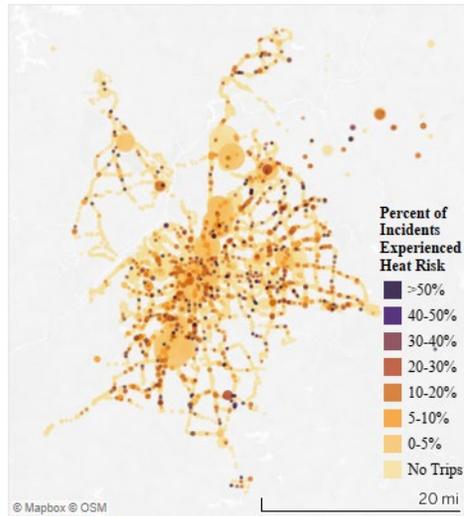 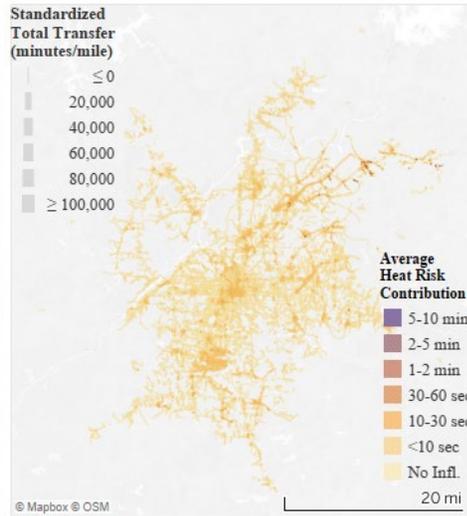 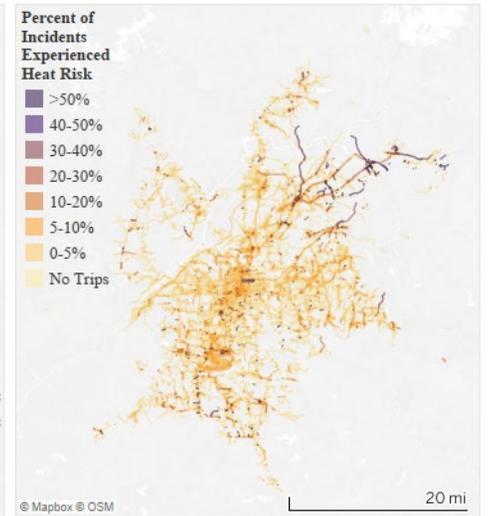
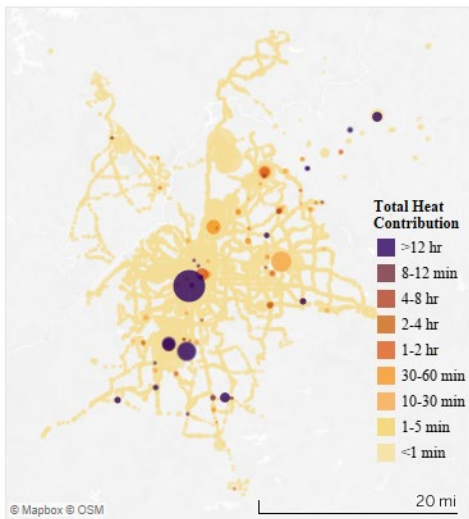 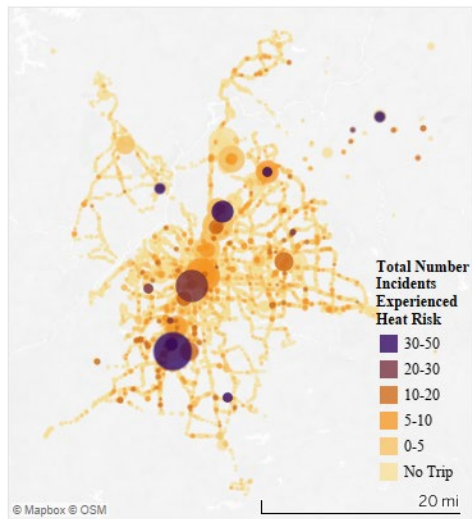 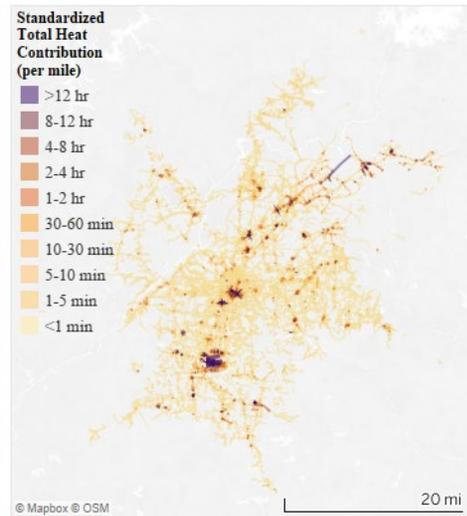 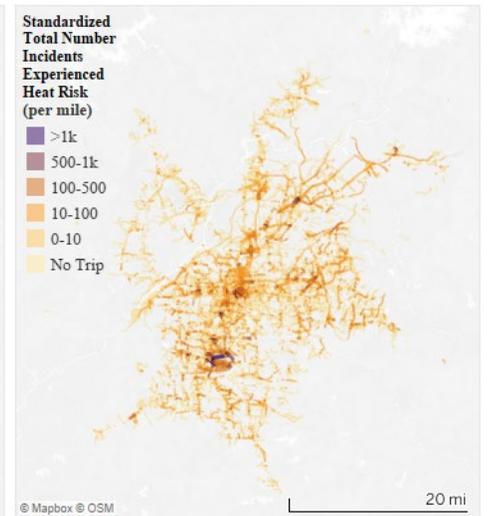

**Figure 6.** Heat Exposure, Vulnerability, and Prioritization

**Comparison with Other Studies**

To investigate differences across recently proposed mesoscale methods, we compared three analytical approaches: (1) the baseline method, which employs the Dijkstra algorithm and additive exposure calculations; (2) the spatio-temporal approach, which uses a spatio-temporal transit routing algorithm and additive exposure calculations; and (3) the current approach, which incorporates a spatio-temporal transit routing algorithm and dynamic cumulative exposure calculations. Additive exposure calculations were adapted from the modeling schemes used by Huang et al. (2024) and L. Liu et al. (2024), employed the NWS' heat categories, and defined factors to quantify the impact of exposure by quantiles. We maintain consistent thresholds between the baseline and Spatio-temporal approach to ensure comparability, but only the pattern of their distribution, not the absolute values, are comparable to the current approach due to the use of different quantifying methods.

Figure 7 shows the overall comparison of travelers' home locations and risk levels for each trip. Compared with the spatiotemporal approach (31.9% of trips under extreme heat risk), the baseline approach (1.7%) has a significant underestimation of heat impact. Under the baseline approach, trips that experience heat risk are often associated with home locations near the areas where Heat Index is high due to the high percentage of impervious surfaces. The spatio-temporal approach, on the other hand, highlights major transit stations. The current approach shows a pattern that highlights the overall system dynamic.

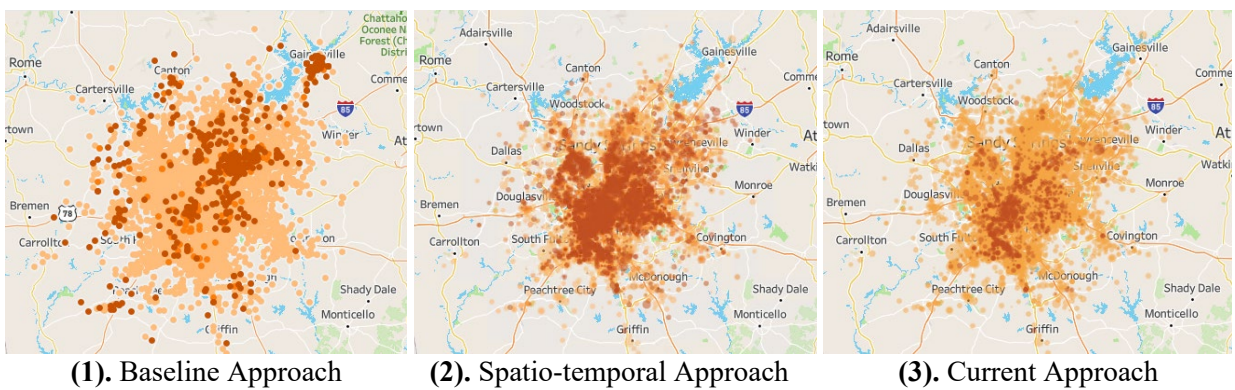

   **(1).** Baseline Approach     **(2).** Spatio-temporal Approach     **(3).** Current Approach

**Figure 7.** Comparison among Three Meso-scale Approaches: Overview Spatial Distribution of Travelers Home Location by Risk Levels

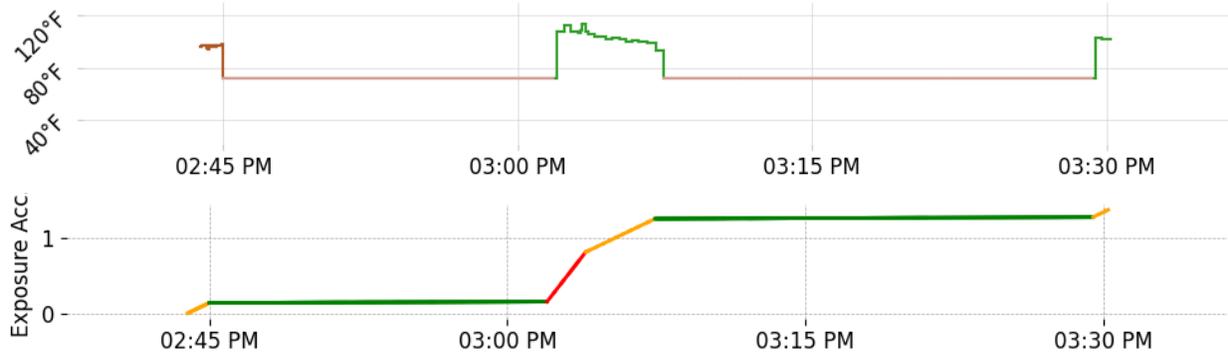

**(1).** Baseline Approach (**Mode colors** – brown: bike, blue: wait, green: walk, pink: transit; **Exposure category colors** – red: dangerous; orange: extreme caution; yellow: caution; green: safe)

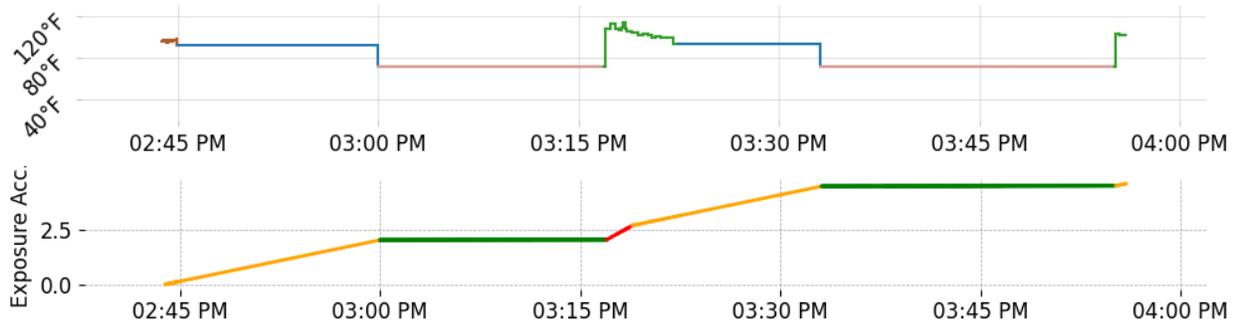

**(2).** Spatio-temporal Approach (**Mode colors** – brown: bike, blue: wait, green: walk, pink: transit; **Exposure category colors** – red: dangerous; orange: extreme caution; yellow: caution; green: safe)

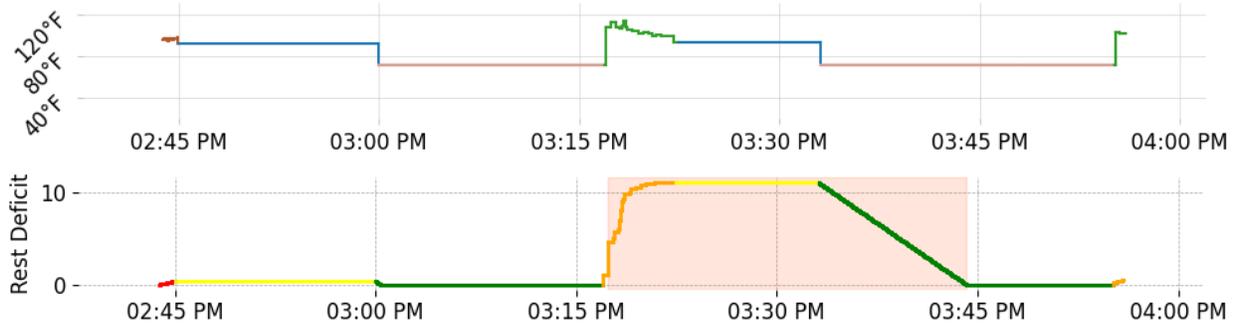

**(3).** Current Approach (**Mode colors** – brown: bike, blue: wait, green: walk, pink: transit; **Physical activities intensity colors** – red: high intensity; orange: moderate intensity; yellow: low intensity; green: rest; coral background: extreme heat flag triggered)

**Figure 8.** Comparison among Three Meso-scale Approaches: Snapshot from One Trip

To better understand the underlying mechanisms of the difference across the three methods, we specifically investigated the example trip presented earlier (Figure 4). The updated temporal trip trajectories for Heat Index variation and intensity levels (physical activity intensity for the current approach and exposure category for the other approaches) are shown in Figure 8. Compared to methods using spatio-temporal algorithms, the base approach (Figure 8-1) neglects the waiting segments of the trip, leading to a significant underestimate of both travel time and heat exposure. Since the exposure scores of the base approach and the spatio-temporal approach were both defined using quantile-based thresholds, their actual health implications are hard to define. The additive and cumulative calculation methods show distinctly different patterns. Comparing the spatio-temporal (Figure 8-2), the current model (Figure 8-1), shows an abrupt increase in exposure (quantified as rest deficit), due to the combined effect of high-temperature exposure and high physical intensity, and a more realistic energy restoration scheme. Overlooking these details could potentially mislead the mitigation-based prioritization into low physical intensity segments (such as the first wait segment), while adaptation-based strategies might incorrectly focus on the final legs of the trip.

## DISCUSSION

### Carryover Effects

The carryover effect, originally conceptualized in pharmacology, describes how the effects of an agent extend beyond its immediate presence. In understanding transit riders' heat exposure, carryover effect can be observed from the discrepancies between mitigation- and adaptation-based prioritization strategies.

Mitigation strategies typically target areas of high physical intensity and thermal discomfort, while adaptation strategies focus on the latter half of the trip when the heat has accumulated significantly. Our findings suggest that while walk ingress and egress contribute similarly to heat accumulation, prioritizing egress over ingress is more effective in terms of adaptation, due to the carryover effect.

This carryover effect also affects asset prioritization. High-exposure areas, identified by mitigation-based strategies, often align with major transportation corridors and densely urbanized areas with extensive impervious surfaces. However, adaptation-based vulnerability scores are more widespread, indicating that travelers accumulate risk in these high-exposure areas and carry this risk into subsequent parts of their journey. Consequently, mitigation-based prioritization scores are predominantly high near airports and busy urban centers, while adaptation-based scores peak around transitional areas like Doraville and Cumberland.

Given the carryover effect, transit planning agencies should clearly define the intended outcomes before implementing strategies. In practice, while providing shade and cooling infrastructure in high-exposure areas can help delay or mitigate heat-related illness, offering medical support based solely on high-exposure areas may not effectively address incidents after they occur.

**Policy Implications**

Extreme heat threatens transit riders in Metro Atlanta, notably during the peak heat hours of 2-4 P.M. With 10.2% of summer transit trips, equating to 20,238 daily, exposed to heat levels surpassing NIOSH's occupational safety standards, the risk is pronounced. Given an average of 1.83 waits per trip at ingress and transfer, and 27.0 minutes of total walking over ingress, egress, and transfer, the risk is particularly apparent during wait and walk segments. Wind chill does not pose a notable threat.

Infrastructure enhancement is needed to combat heat exposure effectively. High-exposure stations such as Southlake and the Airport should expand heat-mitigation infrastructure. Measures could include installing reflective surfaces, increasing canopy coverage, and establishing cooling stations. At high-vulnerability stations like Doraville and Cumberland, installing shaded areas and air-conditioned waiting rooms would enhance comfort and safety. Focuses improvements should prioritize critical transfer stations with extended wait times, such as Five Points Station, Lindbergh Center, West End, and Airport Station.

Adjusting transit schedules to increase frequency during peak heat hours can minimize wait times and reduce heat exposure. This approach is especially critical at stations known for high exposure, such as Southlake and Airport, helping to prevent heat-related illnesses before they develop. In addition, policies should aim to support additional programs that improve conditions for zero-car households, low-income groups, and the elderly, increasing resilience against heat impacts.

In summary, we suggest targeted interventions that not only recognize the spatio-temporal disparity but also the difference between mitigation- and adaptation-based strategies. A detailed policy brief has been developed to summarize these recommendations (Fan, Lyu, Lu, et al., 2024).

**CONCLUSIONS AND RECOMMENDATIONS OFR FURTHER STUDY**

This study presents a novel approach to evaluating transit rider exposure to extreme temperature conditions, by coupling a large dataset of regional onboard transit survey responses with transit schedule/route data and with temperature data. This analytical framework helps fill the current research gap in transportation thermal comfort research by integrating micro-level temperature derivation and macro-level trip analysis and addressing the dynamics in transit trips and cumulative exposure.

Based on the findings of the analysis, extreme heat is a threat to public transit users that needs more attention. In the summer of 2019 (August), 10.2% of all trips are found at risk of extreme heat, with a level of exposure surpassing the occupational safety standards set by NIOSH. Wind chill, however, does not cause any significant threat to observed trips throughout these 10 months. The analysis shows the importance of incorporating resiliency-oriented features such as microclimate improvements designed to mitigate exposure with an increased concern for extreme temperature impacts on transportation and public health.

From transit users' perspective, the study identifies two types of areas around metro Atlanta that should be prioritized for urban heat island mitigation, places with mitigation needs (Southlake, Dawson, and ATL Airport), and locations that require adaptation (Doraville, Cumberland, and Buckhead). While investments in the former support reduced accumulation of heat exposure before a medical condition arises, the latter demands attention to the provision of medical resources for potential heat-related illnesses.

This study utilizes land surface temperature (LST) derived from satellite imagery to extrapolate spatial variations in air temperature. This method effectively captures the mesoscale impacts of urban heat islands and greenery cooling effects, which are not as discernible from meteorological station data alone (Fan et al., 2019). However, LST data present limitations; LST can sometimes overestimate the magnitude of urban heat island effects and exhibits variability near urban greenery (Du et al., 2024; Zhang et al., 2014). Comparisons of satellite-derived LST with ambient air temperature have shown that despite some discrepancies, LST generally aligns well with air temperature patterns, confirming its utility in representing spatial temperature variations (Lokoshchenko et al., 2022; Luo et al., 2018). Additionally, cloud cover significantly influences the accuracy of satellite-based LST products, as high cloud cover can skew LST estimates (Tomlinson et al., 2011). To minimize this effect, this study employed primarily cloud-free images (cloud cover < 10%), thereby reducing potential biases and more closely reflecting actual air temperature conditions.

**ACKNOWLEDGEMENTS**

This study is funded and supported by the National Center for Sustainable Transportation (NCST). The authors also want to thank the Atlanta Regional Commission (ARC) for granting us access to the Transit Onbaord Survey, for their long-term activities associated with the continual improvement of the activity-based travel demand model, and for their never-ending support of research at Georgia Tech by providing access to model-run outputs and their technical expertise.

**APPENDIX A. LAND SURFACE TEMPERATURE RETRIEVAL METHODS**

In deriving LST, cloud masking was conducted based on the Quality Assessment bands (Sayler, 2022). The LST is calculated from blackbody temperature based on Planck's radiation function (Equation 1):

$$T = \frac{k_1}{\ln(\frac{k_2}{T_B}+1)} \qquad (1)$$

Where $k_1$, $k_2$ are pre-defined parameters obtained from the Landsat product metadata, and $T_B$ is the blackbody temperature derived from the thermal radiance ($L_\lambda$) (Equation 2). $L_\lambda$ represents the Top of Atmosphere (TOA) radiance collected at the sensor's level. $T_B$ can be calculated from TOA after accounting for upward radiance ($L_\lambda^\uparrow$) and downward radiance ($L_\lambda^\downarrow$), (Equation 2) with surface emissivity ($\epsilon_\lambda$) and atmospheric transmittance ($\tau_\lambda$) (Prata et al., 1995). This approach has been widely validated and applied in retrieving approximate air temperature for analysis of health and thermal comfort (Efe et al., 2016; Jenerette et al., 2016; Pearsall, 2017; Wilson, 2020).

$$L_\lambda = (\epsilon_\lambda \times T_B \times \tau_\lambda) + (L_\lambda^\uparrow \times \tau_\lambda) + (1 - \epsilon_\lambda) \times L_\lambda^\downarrow \qquad (2)$$

# APPENDIX B. TEMPERATURE EXTRAPOLATION OUTPUT

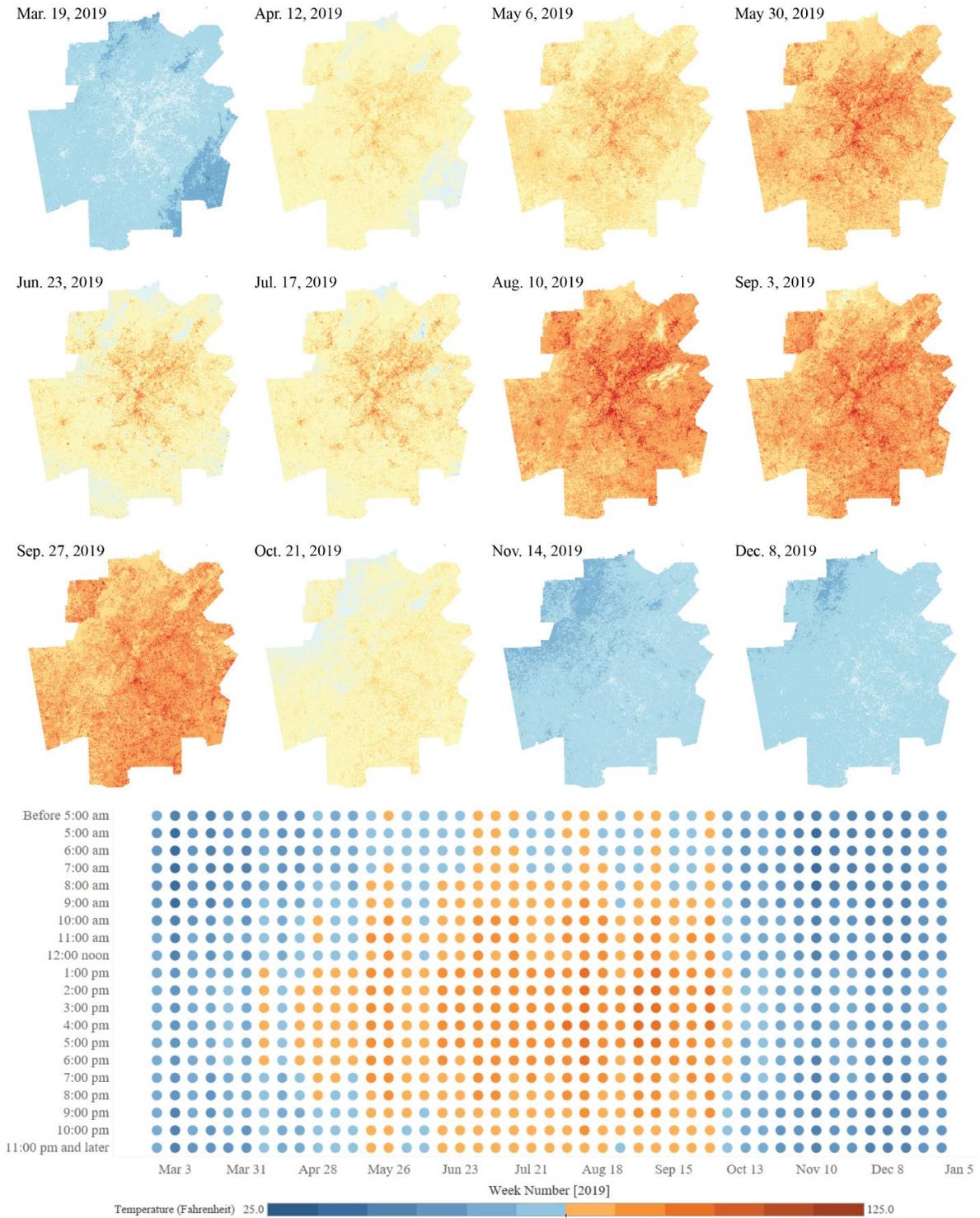

**Figure 9.** Derived Temperature (Top: Over the Study Area every 24 days; Bottom: Average Temperature at the ATL Airport by Day and Hour)

# APPENDIX C. CRITICALITY OF TRANSIT NETWORK ELEMENTS

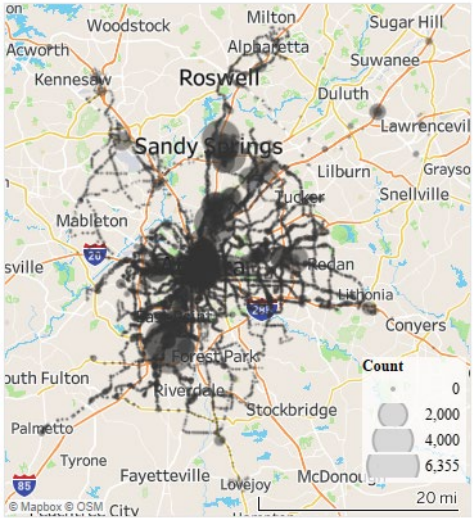
(a). Number of Wait Incidents

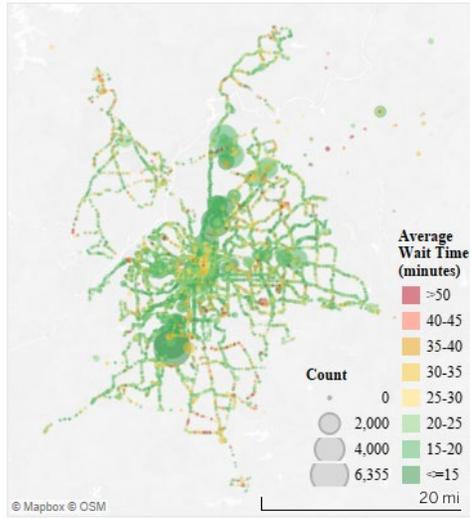
(c). Number of Incidents by Average Wait Time

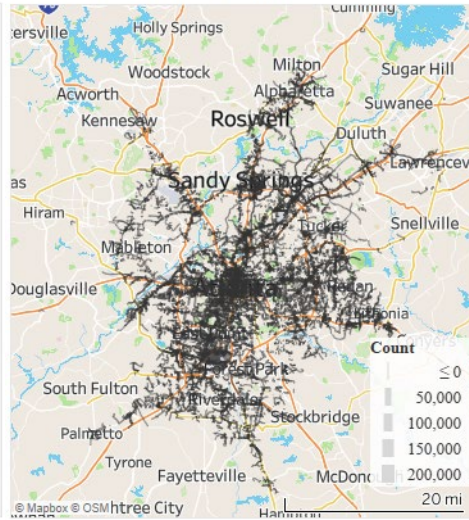
(e). Number of Transfer Incidents

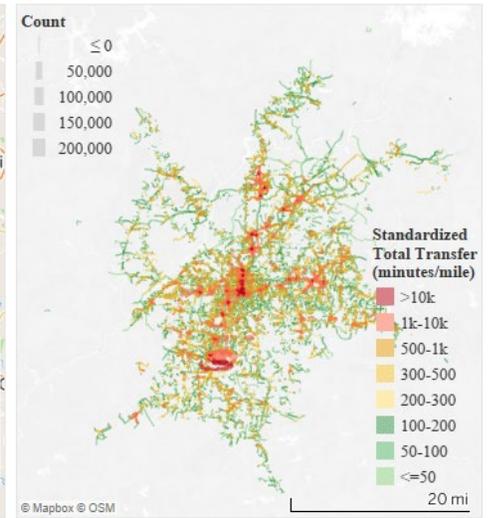
(g). Total Transfer Time by Number of Incidents

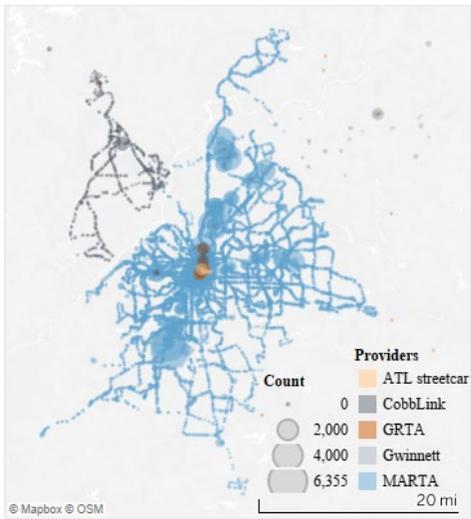
(b). Number of Wait Incidents by Provider

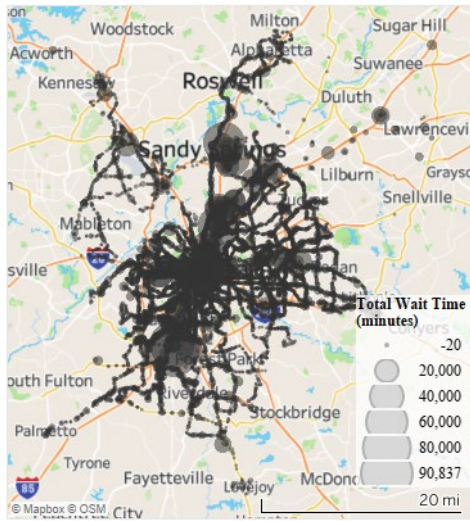
(d). Total Wait Time in Each Station

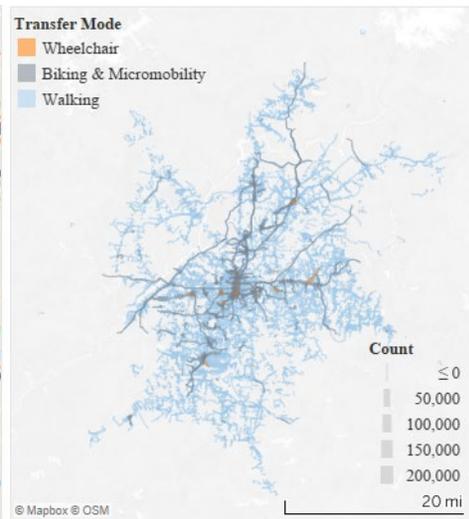
(f). Number of Transfer Incidents by Mode

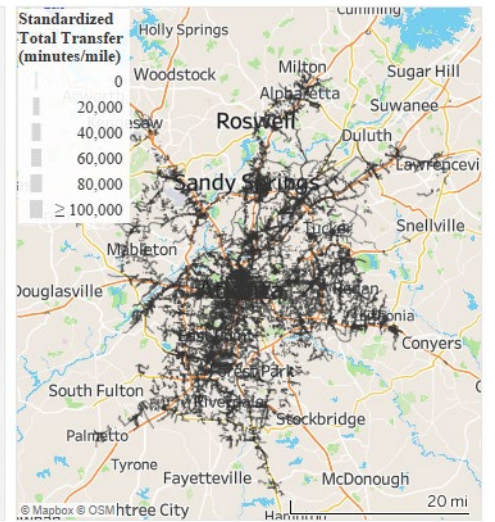
(h). Total Transfer Time in Each Segment

**Figure 10.** Criticality Measures for Transit Stations and Footpaths

# APPENDIX D. EXPOSURE, VULNERABILITY, AND PRIORITIZATION BASED ON WIND CHILL MODELING

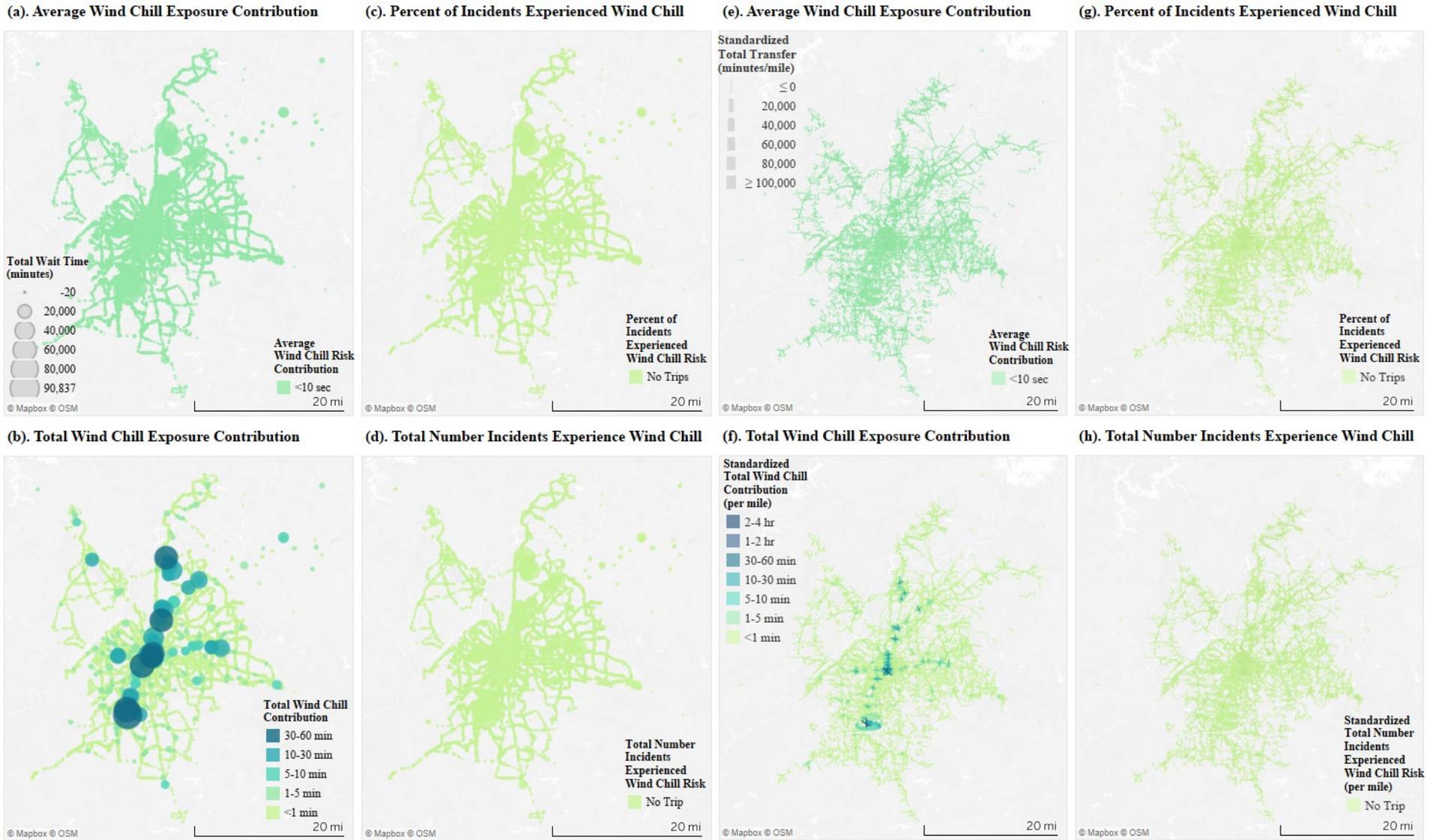

**Figure 11.** Wind Chill Exposure, Vulnerability, and Prioritization